\documentclass[runningheads]{llncs}
\usepackage[T1]{fontenc}
\usepackage{graphicx}
\usepackage{booktabs}

\newcommand{\corr}{\textsuperscript{*}}
\usepackage{mwe}
\usepackage{subfiles}
\usepackage{subcaption}
\usepackage{amsmath,amssymb,amsfonts}
\newcommand{\figpath}{figures}
\usepackage{pgf}
\usepackage{xcolor}
\usepackage{multirow}
\usepackage[hidelinks]{hyperref}
\usepackage{algorithm}
\usepackage{algpseudocode}
\usepackage{import}
\algtext*{EndIf}
\algtext*{EndFor}
\algtext*{EndWhile}
\algtext*{EndFunction}
\algtext*{EndProcedure}
\begin{document}

\title{Sinkhorn Hamiltonian Monte Carlo for Entropic Optimal Transport Generalized Bayes}

\titlerunning{Sinkhorn HMC for Entropic OT Generalized Bayes}
% If the full title of your paper is short enough to also fit in the running head, you can omit the abbreviated paper title here. You can check as follows: if you comment out the \titlerunning line, something will appear in the header of all odd-numbered pages of your PDF from page 3 onward. This something is either the full title (in which case all is well), or the error message "Title Suppressed Due to Excessive Length". If this error message appears, you're going to want to provide an abbreviated title within the \titlerunning command, because if you won't do it, Springer will do it for you.

%N.B.: Author information (both in the \author{} and \authorrunning{} command) should only be present in the Camera-Ready Version of your paper. The version that you initially submit for review, ought to be double-blind. So, when initially submitting your paper, use:
%\author{Author information scrubbed for double-blind reviewing}
\author{Guilhem Nespoulous\inst{1,2}\orcidID{0009-0004-2716-0036} \corr \and
Fr\'ed\'eric Bertrand\inst{1,3}\orcidID{0000-0002-0837-8281}  \and
Myriam Maumy\inst{1,4}\orcidID{0000-0002-4615-1512}\and
Yoann Valero\inst{2}}
% You may leave out the orcidID information, if you want to.
% Use \corr to indicate the corresponding author. Note the spacing around the \corr command. Only one author can be the corresponding author.

%N.B.: comment out the \authorrunning{} command for the double-blind version of your paper submitted for review. Later, if your paper is accepted, use the command for the Camera-Ready Version.
\authorrunning{G. Nespoulous et al.}

\institute{LIST3N, University of Technology of Troyes, Troyes, France\\ \email{guilhem.nespoulous1@utt.fr} % 12 Rue Marie Curie, 10300 
\and
QAD Inc., CA, United States %\email{guilhem.nespoulous@qad.com} % 93108 100 Innovation Place Santa Barbara
\and Cedric, EPN06, Conservatoire national des arts et métiers, Paris, France% 292 Rue Saint-Martin, 75003 
\and Arènes, CNRS 6051, RSMS, EHESP, Rennes, France% 15 Av. du Professeur Léon Bernard, 35043
}
\maketitle

\begin{abstract}
Bayesian posterior sampling is a ubiquitous paradigm for problems where a point estimate of parameters is not sufficient, such as risk analysis and uncertainty quantification. However, likelihoods may be misspecified, intractable, computationally expensive, or not representative of the discrepancy of interest. Generalized Bayes extends likelihood-based posterior updates by using other losses. Sinkhorn divergences have appealing geometric properties: they compare empirical measures directly and yield smooth gradients thanks to entropic regularization. 
In this work, we introduce Sinkhorn divergences as Generalized Bayes losses for Hamiltonian Monte Carlo (HMC) and No-U-Turn Sampler (NUTS). We also propose heuristics to set hyperparameters that affect the stability and calibration quality, such as the number of Sinkhorn iterations, the entropic regularization strength, and the marginal relaxation penalty.  
In regimes where the forward model relies on a stochastic simulator, we combine HMC/NUTS with a common-random-numbers strategy to obtain a deterministic surrogate objective that preserves gradients and Hamiltonian dynamics. We study both mass-preserving balanced and relaxed unbalanced settings. 
We evaluate our method empirically on (1) a simple Gaussian model as a sanity check; (2) a distribution supported on a noisy spiral manifold where a likelihood-based approach is a poor fit; (3) a Gaussian pulse model with misalignment due to errors-in-variables, emphasizing robustness to misspecification; and (4) CIFAR-10 image patch alignment under perturbations, highlighting differences between balanced and unbalanced regimes.

\end{abstract}

\section{Introduction}

Bayesian calibration and posterior inference are based on the principle that we can produce a set of models representative of our posterior knowledge about the credibility of parameter values, given our prior and the available data. In other words, for a given parametric simulator $G_\theta$ we combine our prior knowledge $p(\theta)$ with data $y$, and target the likelihood-based posterior $p(\theta\mid y) \propto p(\theta)\,p(y\mid \theta)$. This method's main advantage over gradient descent is that we do not obtain a point estimate (a single vector) of the targeted parameters, but rather a full distribution \cite{murphy2012machine}. The spread of this posterior distribution is meant to represent the uncertainty about the modeled phenomenon, which may be critical in some real-world situations (risk analysis, etc.). This also enables Monte Carlo assessment of model outcomes under the posterior \cite{robert2004monte}, capturing the effects of parameter uncertainty on further KPIs.

This method's reliance on likelihood evaluations makes it impractical in common real-world situations: likelihoods may not be available in closed form (e.g., implicit simulators), may be computationally heavy to compute, may be misaligned with the final objective, and supports may not match. Furthermore, in practice, models can be misspecified for a wide range of reasons, such as high phenomenon complexity, omission of exogenous effects, measurement imprecision and artifacts, or data drift. All of these can lead to structural model mismatch, implying that no $\theta$ can be considered \textit{true}, as it does not depict the actual phenomenon of interest correctly. In such settings, generally referred to as \emph{M-open} \cite{m-open}, the goal is therefore to find a \textit{useful} posterior set of model parameters, where the meaning of \textit{useful} depends on the study objective and the loss considered.

In this work, we propose to use Sinkhorn divergences as the loss $L$ in Generalized Bayes \cite{BissiriHolmesWalker2016}, where $L(\theta;y)$ quantifies the discrepancy between simulated and observed data, so that $p(\theta \mid y) \ \propto\ p(\theta)\,\exp\!\big(-\eta\,L(\theta;y)\big)$. %attention on peut se poser la question de ce que signifie eta
Sinkhorn divergences directly compare measures and are geometric in the sense that they can handle support mismatch. These divergences are convenient because they are differentiable everywhere and provide built-in hyperparameters to control their smoothness via entropic regularization, which is critical for our targeted gradient-based posterior sampling methods (gradient smoothness w.r.t. model parameters). For this reason, we prefer them to less practical Optimal Transport (OT) quantities like exact Wasserstein. Furthermore, Sinkhorn has balanced and unbalanced variants which enable us to work on a wide range of phenomena. While having such nice properties, we still need to consider potential challenges including simulator determinism, since HMC/NUTS \cite{Neal2011HMC,NUTS} need a deterministic potential. This issue is addressed later by introducing a common-random-numbers averaged surrogate loss. %Ajouter référence OT (Villani) 
\paragraph{Potential applications.} This method could be implemented in a wide variety of applications. Natural examples are geometric phenomena where using OT discrepancies makes particular sense. OT discrepancies are used in diverse domains, like computer vision (discrepancy between images), econometric model calibration (discrepancy over returns). More broadly, the method can be relevant with errors-in-variables and horizontal-noise-prone data, robustness to all kinds of misspecification, physical calibration with noisy data, and others. %Ajouter une référence pour les modèles économétriques

\section{Related Work}

\subsection{Generalized Bayesian inference}

Generalized posteriors, also known as Gibbs posteriors, of the form $p(\theta \mid y) \ \propto\ p(\theta)\,\exp\!\big(-\eta\,L(\theta;y)\big)$ have been widely studied \cite{BissiriHolmesWalker2016,Catoni2007,ChernozhukovHong2003}. They enable application-based losses while preserving uncertainty quantification and leveraging prior knowledge. 

The choice of temperature $\eta$ is discussed in \cite{SyringMartin2019} where the authors calibrate a credible coverage region with bootstrap and solve for $\eta$ via stochastic root-finding. For non-i.i.d. data, calibration can instead be based on frequentist coverage, using truncated samples or multiple datasets when available.% What about non iid - Dire ce qu'est a et b

\subsection{Entropically regularized OT and Sinkhorn divergences}

Development of the Sinkhorn distances \cite{cuturi2013sinkhorn}, and later, divergences \cite{genevay_learning_2017,Feydy} made it possible to avoid solving a constrained linear program directly and instead yield a smooth, convex objective. We consider two discrete measures
$\mu=\sum_i a_i\,\delta_{x_i}$ and $\nu=\sum_j b_j\,\delta_{y_j}$ where $a$ and $b$ denote their marginal weight vectors, and $m_a$ and $m_b$ their total masses. % OT dire que c'est le transport optimal entre deux mesures, 

\begin{equation}
\label{eq:entropic_ot}
\mathrm{OT}_\varepsilon(\mu,\nu)
=\min_{P\in\Pi(a,b)} \Big\{\langle P,C\rangle
+\varepsilon \sum_{i,j} P_{ij}\big(\log P_{ij}-1\big)\Big\}.
\end{equation}

In \cite{genevay_learning_2017}, a debiased expression was introduced to enforce zero discrepancy between two identical measures. % Mettre une formulation avec le X2 instead of 1/2 1/2
\begin{equation}
\label{eq:sinkhorn_div}
S_\varepsilon(\mu,\nu)
=\mathrm{OT}_\varepsilon(\mu,\nu)
-\tfrac12\,\mathrm{OT}_\varepsilon(\mu,\mu)
-\tfrac12\,\mathrm{OT}_\varepsilon(\nu,\nu).
\end{equation}

Unbalanced OT \cite{chizat2017} versions of the Sinkhorn divergences were later introduced in \cite{sejourne2019sinkhornunbalanced} enabling control of mass creation and destruction with a parameter $\rho$ by relaxing the constraints on the marginals. When the total masses are equal, the additional term vanishes and the expression reduces to the usual debiased form.

\begin{equation}
\label{eq:unbalanced_entropic_ot}
\begin{aligned}
\mathrm{OT}^{\mathrm{U}}_{\varepsilon,\rho}(\mu,\nu)
= \min_{P\in\mathbb{R}^{n\times m}_+}\Big\{
&\langle P,C\rangle
+\varepsilon \sum_{i,j} P_{ij}\big(\log P_{ij}-1\big) \\
&+\rho\,\mathbf{KL}(P\mathbf{1}_m\,\|\,a)
+\rho\,\mathbf{KL}(P^\top\mathbf{1}_n\,\|\,b)
\Big\},
\end{aligned}
\end{equation}

\begin{equation}
S^{\mathrm{U}}_{\varepsilon,\rho}(\mu,\nu)
=\mathrm{OT}^{\mathrm{U}}_{\varepsilon,\rho}(\mu,\nu)
-\tfrac12\,\mathrm{OT}^{\mathrm{U}}_{\varepsilon,\rho}(\mu,\mu)
-\tfrac12\,\mathrm{OT}^{\mathrm{U}}_{\varepsilon,\rho}(\nu,\nu)+\frac{\varepsilon}{2}\big(m_a - m_b\big)^2.
\end{equation}

This debiased unbalanced Sinkhorn divergence is the loss we chose to implement in our proposed Generalized Bayesian posterior inference method.

%\subfile{03_ot_nonsmoothness.tex}
%\subfile{04_entropic_sinkhorn.tex}

\section{Proposed method}

\subsection{Generalized Bayes update with Sinkhorn divergences}

Let $\bar S(\theta)$ be an averaged debiased Sinkhorn divergence (see \ref{sec:CRN}) with quadratic cost\footnote{$c(x,y)=\|x-y\|_2^2 \qquad
C_{ij}=c(x_i,y_j).$ } between the observed measure defined in algorithm \ref{alg:sgb-nuts-generic} by $(\mathcal{Y},\log b)=\texttt{build\_measure(}y\texttt{)}$ and the corresponding simulated measure $(\mathcal{X}_\theta,\log a_\theta)=\texttt{build\_measure(}G(\theta,\omega)\texttt{)}$. 
Let $p(\theta\mid y)$ be defined as a generalized posterior: 
\begin{equation}
p(\theta\mid y)\ \propto\ p(\theta)\,\exp\!\big(-\eta\,\bar S(\theta)\big).
\end{equation}
If $G$ is deterministic, as explained in \ref{sec:CRN}, we have $K=1$, which removes averaging. Algorithms \ref{alg:sgb-nuts-generic} and \ref{alg:debiased-sinkhorn} introduce the detailed procedure that we implemented. % Dire que G est un si

\paragraph{Measure construction.} The function \texttt{build\_measure()} outputs a weighted measure with support points $\mathcal{X}=\{x_i\}_{i=1}^n\subset\mathbb{R}^d$ and corresponding logarithmic weights \texttt{log a}. With point cloud data (experiments \hyperref[sec:expA_normal]{A}, \hyperref[sec:expB1_spiral]{B}, \hyperref[sec:expC]{C}), $\mathcal{X}$ corresponds to the sampled locations in the data space. In \hyperref[sec:expD]{Experiment D} the support is fixed and corresponds to a pixel grid (for both observation and simulation). 

The weights can be (1) uniform, with $a_i=\frac{1}{n}$ and $\log a_i=-\log n$; (2) from raw masses, $a_i=w_i, \log a_i=\log w_i$, as in the unbalanced version of \hyperref[sec:expD]{experiment D} where $w_i$ is the pixel intensity; or (3) from normalized masses as in the balanced version of \hyperref[sec:expD]{experiment D} so that the total mass is fixed:
\begin{equation}
a_i = \frac{w_i}{\sum_{\ell=1}^{n} w_\ell},
\qquad
\log a_i = \log w_i - \log\!\Big(\sum_{\ell=1}^{n} w_\ell\Big).
\end{equation}
Balanced Sinkhorn requires equal total masses ($\sum_{i=1}^{n} a_i=\sum_{j=1}^{m} b_j$).

\paragraph{Loss scale correction.}\label{par:loss-corr} If the measure construction is normalized, a natural choice is to scale with the number of support points (e.g. number of pixels $HW$) so \texttt{loss\_obs\_scale()} returns $n$. If weights are not normalized, it returns $1$ so the Generalized Bayes updates stay comparable in scale across experiments.

\begin{algorithm}[!htbp]
\caption{Sinkhorn divergences as Generalized Bayes loss for Hamiltonian Monte Carlo/NUTS sampling.}
\label{alg:sgb-nuts-generic}
\begin{algorithmic}[1]
\Require Prior $p(\theta)$, observations $y$, simulator $G(\theta,\omega)$
\Require \textsc{build\_measure}$()$: data and simulation output to $(\mathcal{X},\log a)$
\Require OT settings: cost $c$, $\varepsilon$, $\rho$ (if unbalanced), $T$ iters, flag \texttt{balanced}%, \texttt{include\_yy}
\Require Loss scale $\lambda$, \textsc{loss\_obs\_scale}$()$ the function used for loss scaling, CRN replicates $K$ if $G$ is not deterministic

\State $(\mathcal{Y},\log b)\gets \textsc{build\_measure}(y)$ 
\State Choose $\varepsilon$ (e.g.\ $\varepsilon=(\alpha h)^2$ from support resolution)
\If{$G$ is stochastic}
  \State Draw CRN seeds $\omega_{1:K}$ \Comment{deterministic surrogate objective in $\theta$}
\Else
  \State $K\gets 1$
\EndIf

\Function{Loss}{$\theta$}
  \For{$k=1,\dots,K$}
    \State $s^{(k)} \gets G(\theta,\omega_k)$
    \State $(\mathcal{X}^{(k)}_\theta,\log a^{(k)}_\theta)\gets \textsc{build\_measure}(s^{(k)})$
    \State $S_k \gets \textsc{debiased\_sinkhorn}\big((\mathcal{X}^{(k)}_\theta,\log a^{(k)}_\theta),(\mathcal{Y},\log b),\varepsilon, \rho,T,\texttt{balanced}\big)$%,\texttt{include\_yy}
  \EndFor
  \State \Return $\bar{S}(\theta)\gets \frac{1}{K}\sum_{k=1}^K S_k$
\EndFunction

\State $\eta \gets \lambda \times \textsc{loss\_obs\_scale}(\mathcal{Y},\log b)$ 
\State Target generalized posterior: $\,p(\theta\mid y)\propto p(\theta)\exp\!\big(-\eta\,\bar{S}(\theta)\big)$
\State Run NUTS on $U(\theta)= -\log p(\theta)+\eta\,\bar{S}(\theta)$ (autodiff through $G$ and fixed-$T$ Sinkhorn)
\end{algorithmic}
\end{algorithm}

\begin{algorithm}[!htbp]
\caption{\textsc{debiased\_sinkhorn}$()$: Debiased Sinkhorn divergences (balanced or mass-relaxed)}
\label{alg:debiased-sinkhorn}
\begin{algorithmic}[1]
\Require Weighted measures $(\mathcal{X},\log a)$ and $(\mathcal{Y},\log b)$, cost $c$, $\varepsilon$, $\rho$, iters $T$, flag \texttt{balanced}%, \texttt{include\_yy}

\Function{\textsc{sinkhorn\_OT\_value}}{$C,\log a,\log b,\varepsilon,\rho,T,\texttt{balanced}$}
  \State $\tau \gets 1$ if \texttt{balanced} else $\rho/(\rho+\varepsilon)$
  \State Initialize $f\gets 0\in\mathbb{R}^n$, $g\gets 0\in\mathbb{R}^m$ \Comment{dual potentials}
  \For{$t=1,\dots,T$}
    \State $f_i \gets \tau\Big(-\varepsilon\log\sum_{j=1}^m \exp\big((g_j-C_{ij})/\varepsilon + \log b_j\big)\Big)\ \forall i$
    \State $g_j \gets \tau\Big(-\varepsilon\log\sum_{i=1}^n \exp\big((f_i-C_{ij})/\varepsilon + \log a_i\big)\Big)\ \forall j$
  \EndFor
    \State \Return $OT_{\varepsilon}(a,b)$ if \texttt{balanced}, else $OT^{\mathrm U}_{\varepsilon,\rho}(a,b)$, evaluated at $(f,g)$
\EndFunction

\State $C_{XY}[i,j]\gets c(x_i,y_j)$, \quad $C_{XX}[i,i']\gets c(x_i,x_{i'})$, \quad $C_{YY}[j,j']\gets c(y_j,y_{j'})$
\State $OT_{XY}\gets \textsc{sinkhorn\_OT\_value}(C_{XY},\log a,\log b,\varepsilon,\rho,T,\texttt{balanced})$
\State $OT_{XX}\gets \textsc{sinkhorn\_OT\_value}(C_{XX},\log a,\log a,\varepsilon,\rho,T,\texttt{balanced})$
\State $OT_{YY}\gets \textsc{sinkhorn\_OT\_value}(C_{YY},\log b,\log b,\varepsilon,\rho,T,\texttt{balanced})$
\State $m_a \gets \sum_{i=1}^n \exp(\log a_i)$, \quad $m_b \gets \sum_{j=1}^m \exp(\log b_j)$
%\If{\texttt{include\_yy}}

\State \Return $OT_{XY}-\tfrac12 OT_{XX}-\tfrac12 OT_{YY}+\tfrac{\varepsilon}{2}(m_a-m_b)^2$
%\Else
  %\State \Return $OT_{XY}-\tfrac12 %OT_{XX}+\tfrac{\varepsilon}{2}(m_a-m_b)^2$ %\Comment{no observed data debiasing}
%\EndIf
\end{algorithmic}
\end{algorithm}

\subsection{Common-random-numbers (CRN) for stochastic simulators}\label{sec:CRN}

We use the CRN-averaged surrogate loss when our simulators are stochastic. The simulator randomness can include latent time grids, Gaussian noise arrays, discrete event variables and any other random draws used by the simulator $G$. The randomness is denoted by $\omega \sim r(\omega)$ where $r(\omega)$ is the distribution of that randomness. In such a situation, the ideal loss $S^{\star}(\theta)$ is obtained using the expectation of the Sinkhorn divergence with respect to $\omega$. 
\begin{equation}
S^{\star}(\theta)
=
\mathbb{E}_{\omega}\!\left[\,S(\theta,\omega)\,\right],
\qquad
p^{\star}(\theta\mid y)  \propto p(\theta)\,\exp\!\big(-\eta\,S^{\star}(\theta)\big). \end{equation}
Generally, this is not practically tractable. To bypass this, we first draw $\omega_{1:K} \sim r^{\otimes K}$, then we define:

\begin{equation}
\bar S_K(\theta;\omega_{1:K})=\frac{1}{K}\sum_{k=1}^K S(\theta,\omega_k).
\end{equation}
This expression converges\footnote{If $\mathbb E_r\!\left[\,|S(\theta,\omega)|\,\right] < \infty$ for a given $\theta$, $\bar S_K(\theta;\omega_{1:K}) \xrightarrow[K\to\infty]{\text{a.s. under }r^{\otimes K}}
S^\star(\theta)$} to $S^\star(\theta)$ as $K \to \infty$. If $\mathrm{Var}_r\!\left[S(\theta,\omega)\right] < \infty$,
\begin{equation}
\mathrm{Var}\!\left[\bar S_K(\theta;\omega_{1:K})\right]
=
\frac{1}{K}\,\mathrm{Var}_r\!\left[S(\theta,\omega)\right].
\end{equation}
We can define a useful variance estimator as follows:
\begin{equation}
\label{eq:var_estimator}
\widehat{\mathrm{Var}}\!\left[\bar S_K(\theta;\omega_{1:K})\right]
=
\frac{1}{K(K-1)}
\sum_{k=1}^K
\Big(S(\theta,\omega_k)-\bar S_K(\theta;\omega_{1:K})\Big)^2.
\end{equation}
$\bar S_K(\theta;\omega_{1:K})$ is deterministic w.r.t. $\theta$ when $\omega_{1:K}$ is fixed. Therefore, the potential $U_K(\theta)$ constructed from it is also deterministic:
\begin{equation}
U_K(\theta)
=
-\log p(\theta) + \eta\,\bar S_K(\theta;\omega_{1:K}).
\end{equation}
Hence, we can use the following CRN-averaged surrogate loss for HMC/NUTS as our Generalized Bayes updating loss:
\begin{equation}
p_K(\theta\mid y,\omega_{1:K}) \propto p(\theta)\exp\!\big(-\eta\,\bar S_K(\theta;\omega_{1:K})\big).
\end{equation}

\subsection{Selection of loss hyperparameters}
\label{sec:parameter-select}
\paragraph{Loss scale $\lambda$.} We base our approach on \cite{SyringMartin2019}. We select a loss scale parameter $\lambda$ instead of $\eta$ directly, see \ref{par:loss-corr}. Depending on our experimental conditions:
\begin{enumerate}
    \item We know the true parameters. We compute the frequentist coverage of the true parameter using high posterior density (HPD) or Gaussian approximation credible ellipse regions. The regions are computed from posterior samples obtained from multiple datasets generated at the true parameter values. Root-finding or grid search is used to select the highest loss scale that meets the expected coverage.
    \item We do not know the true parameters. We start by using an estimator of the true parameter, in our case we use an approximate MAP estimate $\hat \theta$ obtained from a default loss scale calibration and choosing the posterior sample with the highest probability. We compute the frequentist coverage of $\hat \theta$, then use nonparametric bootstrapping (on i.i.d. data). See \ref{sec:expB1_spiral}.
\end{enumerate}
\paragraph{Marginal relaxation regime and penalty $\rho$.} 
The application mismatch scale should guide the choice of $\rho$. With quadratic cost, $\sqrt{\rho}$ can be interpreted as an approximate displacement scale on the support beyond which mass variation becomes increasingly competitive compared to transport. A very large $\rho$ yields behavior close to the balanced regime, see \ref{sec:expA_normal}.

%\paragraph{Inclusion of the observation debiasing term.} The observation self transportation cost is a constraint used to enforce $S_\varepsilon(\mu,\nu) = 0$, but it is a constant so removing it only changes the objective up to a constant. That said, we include it (\texttt{include\_{yy} = True}) in our experiments as it can be precomputed once and saved.
\paragraph{Common-random-number replicates $K$.}  Picking $K$ is a tradeoff between computation time, bias, and loss geometry. We monitor the computation time increase per replicate and sample, and use the variance estimator in (\ref{eq:var_estimator}) to choose $K$. 

\paragraph{Entropic regularization parameter $\varepsilon$.} We do not set the $\varepsilon$ parameter\footnote{Not to be confused with the HMC leapfrog step size, also commonly denoted by $\varepsilon$.} using annealing methods \cite{epsilon-annealing} as this would deform the loss, threatening the stability of HMC potentials. Setting $\varepsilon$ too low tends to lead to an unstable loss surface with high curvature zones (where the transport plan changes quickly w.r.t. the parameters). Setting it high trades its geometry for increased smoothness. In such regimes the loss behaves increasingly like Maximum Mean Discrepancy counterparts \cite{Feydy}. In our work we choose to use the data resolution heuristic.

Considering that we work with quadratic cost, we set this baseline as $\varepsilon = (\alpha h)^2$ where $\alpha$ is a scaled entropic regularization knob and $h$ is the data resolution. Extending the idea proposed in \cite{medina2025}, we propose to set $h$ as either (1) the measurement noise if known; (2) the observed points' median nearest-neighbor distance if the noise is unknown and we are working with a point cloud; or (3) in case of binned or gridded observations, the smallest bin width or spacing as proposed in \cite{medina2025}. In case of instabilities (trace divergences, poor mixing) $\alpha$ can be increased from a base proposed value of $1$ used throughout this work.

\paragraph{Number of Sinkhorn iterations $T$.}
Budget $T$ impacts on the steps computation complexity. Lower $\varepsilon$ requires more Sinkhorn iterations to converge. We did not thoroughly study the effects of choosing $T$ too low, but in these regimes it seems to introduce some regularization. We propose to monitor $\Delta_S$ and $\Delta_{\nabla}$ which respectively target the loss and gradient difference at representative values of $\theta$:

\begin{equation}
\Delta_S(\theta;T,T_{\mathrm{ref}})
=
\left|\bar S_T(\theta)-\bar S_{T_{\mathrm{ref}}}(\theta)\right|;
\quad
\Delta_{\nabla}(\theta;T,T_{\mathrm{ref}})
=
\frac{
\left\|\nabla \bar S_T(\theta)-\nabla \bar S_{T_{\mathrm{ref}}}(\theta)\right\|
}{
\max\!\left(1,\left\|\nabla \bar S_{T_{\mathrm{ref}}}(\theta)\right\|\right)
}.
\end{equation}

%\subfile{06_implementation_stan.tex}

\newcommand{\expBOnefig}{\figpath/expB1}

\section{Empirical Study and Results}

\subsection{Experimental Setup}
The following experiments were run on a laptop, with an AMD Ryzen 7 5800H CPU, 16 GB RAM, and an NVIDIA RTX 3050 GPU, using Python 3.11, JAX 0.9.0, NumPyro 0.19.0, and Ubuntu 22.04. No heavy tasks were run during calibrations to keep time reports representative; however, light processes may still have some low-level influence on performance. JAX was run on the GPU, and relevant objects are in \texttt{dtype=32}. We sample chains sequentially.

%\begin{table}[ht]
%\centering

%\begin{minipage}[t]{0.48\linewidth}
%\centering
%{\bfseries Hardware}\par\vspace{0.3em}
%{\ttfamily\small
%\begin{tabular}{@{}ll@{}}
%\toprule
%CPU  & AMD Ryzen 7 5800H \\
%CPUs & 16\\
%RAM  & 16 GB \\
%GPU  & NVIDIA GeForce RTX 3050 \\
%\bottomrule
%\end{tabular}
%}
%\end{minipage}\hfill
%\begin{minipage}[t]{0.48\linewidth}
%\centering
%{\bfseries Software}\par\vspace{0.3em}
%{\ttfamily\small
%\begin{tabular}{@{}ll@{}}
%\toprule
%python  & 3.11 \\
%JAX     & 0.9.0 \\
%numpyro & 0.19.0 \\
%Ubuntu  & 22.04 \\
%\bottomrule
%\end{tabular}
%}
%\end{minipage}
%\label{tab:env_summary}
%\end{table}

\subsection{Experiment A: Sanity check with a well-specified normal model}
\label{sec:expA_normal}

\newcommand{\expAfig}{\figpath/expA}

\subsubsection{Experiment description.}

In our first experiment, to check whether our Sink\-horn HMC and CRN pipeline works, we choose a normal model $y_i \sim \mathcal{N}(\mu,\sigma^2)$ with no misspecification. We aim to verify that (1) no numerical or computational pathology occur; (2) the sampling process is stable and meets HMC/NUTS convergence and mixing criteria; (3) the resulting posterior distributions are reasonable (centered close to the true values and with sensible posterior dispersion). We compare our approach to the likelihood HMC/NUTS baseline. As a secondary sanity check, we also run an unbalanced calibration with a very large mass-variation penalty set to $\rho = 10^{6}$ and expect it to behave similarly to the balanced regime. 

\paragraph{Model, data and choice of priors.}

We set true parameters to $\mu^\star=2.0$ and $\sigma^\star=0.5$. We sample $120$ observations from the true model and use priors $ \mu \sim \mathcal{N}(0,5^2)$, $\sigma \sim \mathrm{HalfNormal}(2)$, with the same prior specification for all three settings.

\paragraph{OT loss settings.}
Sinkhorn iterations are set to $80$, we use $K=6$ CRN replicates, $\alpha$ and the loss scale are both set to $1$, yielding $\varepsilon = 2.3853\times 10^{-5}$. We do not sweep through loss hyperparameters, keeping them at a \textit{default} value, as this will be done in further experiments, see \ref{sec:expB1_spiral}.

\paragraph{HMC/NUTS selected calibration parameters.}
We sample $4$ chains each with $500$ warmup and $300$ retained samples. In this experiment, we do not tweak any other NUTS parameters, they are left to NumPyro default values.

\subsubsection{Results and interpretations.}
\begin{figure}[t]
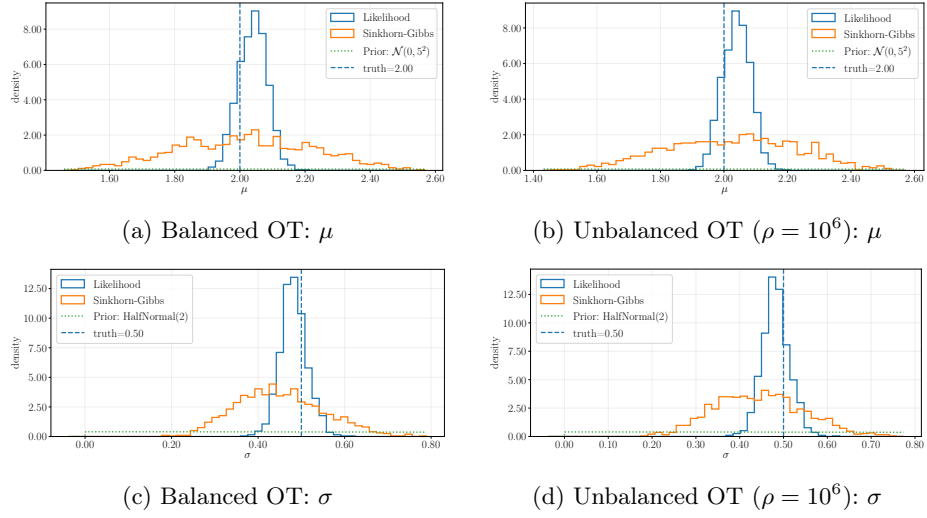

\centering

% ---- MU ----
\begin{subfigure}[t]{0.48\linewidth}
  \centering
  \input{posterior_mu__balanced1_alpha1.0.tex}
  \caption{Balanced OT: $\mu$}
\end{subfigure}\hfill
\begin{subfigure}[t]{0.48\linewidth}
  \centering
  \input{posterior_mu__balanced0_alpha1.0_rho1000000.0.tex}
  \caption{Unbalanced OT ($\rho=10^6$): $\mu$}
\end{subfigure}

\vspace{0.6em}

% ---- SIGMA ----
\begin{subfigure}[t]{0.48\linewidth}
  \centering
  \input{posterior_sigma__balanced1_alpha1.0.tex}
  \caption{Balanced OT: $\sigma$}
\end{subfigure}\hfill
\begin{subfigure}[t]{0.48\linewidth}
  \centering
  \input{posterior_sigma__balanced0_alpha1.0_rho1000000.0.tex}
  \caption{Unbalanced OT ($\rho=10^6$): $\sigma$}
\end{subfigure}

\caption{Experiment A: posterior overlays with likelihood baseline, balanced Sinkhorn, and unbalanced (very large $\rho$).}
\label{fig:expA_normal_posteriors_overlay}
\end{figure}

\begin{figure}[t]
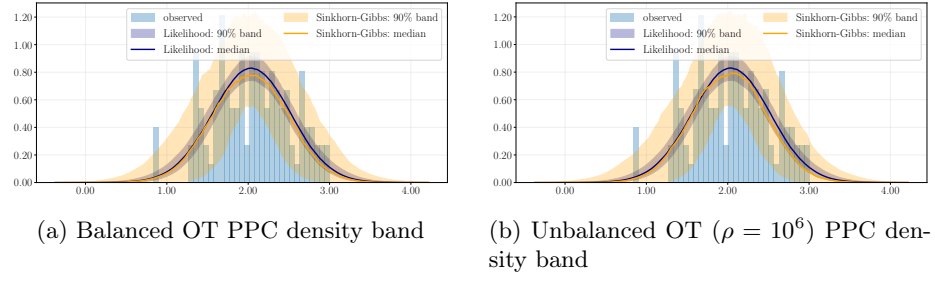

\centering
\begin{subfigure}[t]{0.48\linewidth}
  \centering
  \input{ppc_density_bands__balanced1_alpha1.0.tex}
  \caption{Balanced OT PPC density band}
\end{subfigure}\hfill
\begin{subfigure}[t]{0.48\linewidth}
  \centering
  \input{ppc_density_bands__balanced0_alpha1.0_rho1000000.0.tex}
  \caption{Unbalanced OT ($\rho=10^6$) PPC density band}
\end{subfigure}
\caption{Experiment A: posterior predictive density bands with observed histogram and predictive densities.}
\label{fig:expA_normal_ppc_density_bands}
\end{figure}

As we can see in Table \ref{tab:expA_normal_summary}, (1) The sampling runs without critical errors, at a rate of approximately $24$ samples per second. (2) We do not obtain any divergences, traces mix well, $\hat{R}$ values indicate good convergence. (3) The posterior parameter samples are centered as expected and posterior predictive plots show no obvious problem, see Figure \ref{fig:expA_normal_posteriors_overlay} and \ref{fig:expA_normal_ppc_density_bands}.
These checks also shed light on the fact that the default loss scale seems too low. We would like the posterior to be more concentrated around the true values. The proposed ideas (see \ref{sec:parameter-select}) to select a relevant loss scale w.r.t. the frequentist coverage are implemented and in experiment B, see \ref{sec:expB1_spiral}.

\begin{table}[ht]
\caption{Experiment A: likelihood baseline (\textit{Lkl}), balanced Sinkhorn (\textit{Sink b}) and unbalanced version (\textit{Sink u}) with high mass-variation penalty $\rho=10^6$. }\label{tab:expA_normal_summary}
\centering
\begin{tabular}{lcccccc}
\toprule
Method &
$\mu$ mean (sd) [5\%,95\%] &
$\sigma$ mean (sd) [5\%,95\%] &
$n_{\text{eff}}(\mu)$ &
$n_{\text{eff}}(\sigma)$ &
$\hat R_{\max}$\\
\midrule
Lkl &
2.040 (0.044) [1.964, 2.105] &
0.483 (0.031) [0.435, 0.536] &
1186 &
950 &
1.001 \\
Sink b &
2.015 (0.209) [1.676, 2.359] &
0.450 (0.100) [0.290, 0.613] &
745 &
847 &
0.9999 \\
Sink u &
2.022 (0.206) [1.694, 2.348] &
0.445 (0.102) [0.284, 0.609] &
730 &
767 &
1.008 \\
\bottomrule
\end{tabular}
\end{table}

\providecommand{\expAfig}{./expA_outputs/20260226_135034__normal}

\newcommand{\includepgf}[1]{%
  \IfFileExists{#1}{%
    \resizebox{\linewidth}{!}{\input{#1}}%
  }{%
    \fbox{\ttfamily Missing file: #1}%
  }%
}

% ============================================================
% Posterior overlays (Likelihood vs Sinkhorn) — per Sinkhorn config
% ============================================================

% ============================================================
% Posterior predictive checks (Likelihood vs Sinkhorn) — per Sinkhorn config
% ============================================================

%\begin{figure}[ht]
%\centering

% ---- MU traces ----
%\begin{subfigure}[t]{0.32\linewidth}
%  \centering
%  \includepgf{trace_mu_likelihood.pgf}
%  \caption{Likelihood: $\mu$}
%\end{subfigure}\hfill
%\begin{subfigure}[t]{0.32\linewidth}
%  \centering
%  \includepgf{trace_mu_sinkhorn__balanced1_alpha1.0.pgf}
%  \caption{Balanced OT: $\mu$}
%\end{subfigure}\hfill
%\begin{subfigure}[t]{0.32\linewidth}
%  \centering
%  \includepgf{trace_mu_sinkhorn__balanced0_alpha1.0_rho1000000.0.pgf}
%  \caption{Unbalanced OT ($\rho=10^6$): $\mu$}
%\end{subfigure}
%
%\vspace{0.6em}

% ---- SIGMA traces ----
%\begin{subfigure}[t]{0.32\linewidth}
%  \centering
%  \includepgf{trace_sigma_likelihood.pgf}
%  \caption{Likelihood: $\sigma$}
%\end{subfigure}\hfill
%\begin{subfigure}[t]{0.32\linewidth}
%  \centering
%  \includepgf{trace_sigma_sinkhorn__balanced1_alpha1.0.pgf}
%  \caption{Balanced OT: $\sigma$}
%\end{subfigure}\hfill
%\begin{subfigure}[t]{0.32\linewidth}
%  \centering
%  \includepgf{trace_sigma_sinkhorn__balanced0_alpha1.0_rho1000000.0.pgf}
%  \caption{Unbalanced OT ($\rho=10^6$): $\sigma$}
%\end{subfigure}
%
%\caption{Experiment A: trace plots likelihood baseline, balanced Sinkhorn, unbalanced Sinkhorn (very large $\rho$).}
%\label{fig:expA_normal_traces}
%\end{figure}

\subsection{Experiment B: Calibration on a noisy spiral}
\label{sec:expB1_spiral}

\subsubsection{Experiment description.}
In this experiment, we want to verify that our method is able to produce sensible results in cases where likelihood-based losses are inadequate e.g. multidimensional point clouds and manifold-supported distribution. We consider a spiral dataset where likelihood-based formulations would need to rely on latent-variable modeling. We also cover the loss scale calibration under a (noisy) dataset scenario where the true parameters are unknown.

\paragraph{Model, data and choice of priors.} The spiral is generated from latent positions $t_i \sim \mathrm{Uniform}(0,1)$, adding noise in the form of a $2D$ Gaussian vector $\mathbf{z}_i \sim \mathcal{N}(\mathbf{0}, \textbf{I}_2)$. The $N=160$ observed $2D$ points are then generated as:
\begin{equation}
    \varphi_i = 2\pi \times \mathrm{turns} \times  t_i;\qquad
    c_i = c_0 + \kappa\,\varphi_i;\qquad
    \mathbf{y}_i = \bigl(c_i\cos\varphi_i,\; c_i\sin\varphi_i\bigr) + \sigma \mathbf{z}_i.
\end{equation}
The true parameters are set to $c_0^\star=0.30$, $\kappa^\star=0.15$, $\sigma^\star=0.08$, $\mathrm{turns}^\star=3$. The parameters inferred by the model are the spiral offset $c_0$ and the spiral growth $\kappa$, with respective priors $\mathrm{HalfNormal}(1.0)$ and $\mathrm{HalfNormal}(0.5)$. 

\begin{figure}[ht]
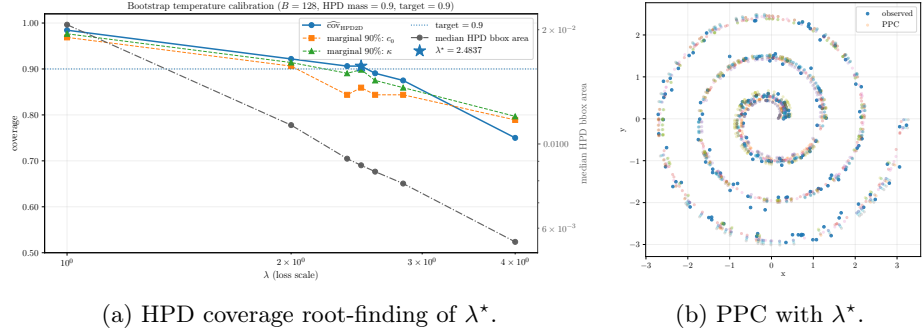

\centering
\begin{subfigure}[t]{0.64\linewidth}
  \centering
  \input{temp-calibration-dualaxis-fixed.tex}
  \caption{HPD coverage root-finding of $\lambda^\star$.}
\label{fig:b2}
\end{subfigure}\hfill
\begin{subfigure}[t]{0.34\linewidth}
  \centering
  \input{ppc\_lambda\_star.tex}
  \caption{PPC with $\lambda^\star$.}
\end{subfigure}
\caption{Experiment B, $\lambda^\star$ selection and PPC from random samples (all chains).}
\label{fig:b}
\end{figure}

\paragraph{OT loss settings.} Same as in Experiment A, except for the choice of the loss scale that is based on Subsection \ref{sec:parameter-select} and  \cite{SyringMartin2019} to better match the frequentist coverage. We start by running the calibration with a default $\lambda_0 = 1$ to obtain $\hat \theta$, an approximate MAP estimate from the $S$ samples\footnote{$\hat\theta \in \arg\max_{1\le s\le S}\ \Big\{\log p(\theta^{(s)}) - \eta_{\lambda_0} \bar S_K(\theta^{(s)})\Big\}.$ Definition of $\eta$ in Alg.\ref{alg:sgb-nuts-generic}.}. Then we create a set of $B=128$ bootstrap datasets $\{y^{*(b)}\}_{b=1}^{B}$. We compute the frequentist coverage of $\hat \theta$ from a $2D$ HPD credible region of samples from each of the bootstrapped datasets, see Figure \ref{fig:b}. Let us denote by $q_b(\lambda)$ the HPD threshold: \begin{equation}   \widehat{\mathrm{cov}}_{\mathrm{HPD2D}}(\lambda)=\frac{1}{B}\sum_{b=1}^{B} \mathbf{1}\!\left\{\log p(\hat \theta) - \eta_{\lambda} \bar S_K(\hat \theta; y^{*(b)}) \ge q_b(\lambda)\right\}. \end{equation}
Instead of recomputing the unnormalized log posterior, we load NumPyro’s potential energy $U(\tilde z)$, and apply a Jacobian correction\footnote{$\log p(\theta^{(s)})-\eta_\lambda \bar S_K(\theta^{(s)};y)
\ \propto\
-\,U\!\left(\tilde z^{(s)};y\right)\;-\;\log\!\left|\det J_T\!\left(\tilde z^{(s)}\right)\right|.$}. As proposed in \cite{SyringMartin2019} we root-find $\lambda^\star$, the highest evaluated temperature such that $\widehat{\mathrm{cov}}_{\mathrm{HPD2D}}(\lambda^\star)\ge 0.9$, evaluating at most 12 temperatures until the precision threshold is met.

\paragraph{HMC/NUTS selected calibration parameters.} Unchanged, see \ref{sec:expA_normal}.

\newcommand{\expBfig}{\figpath/expB}

\subsubsection{Results and interpretations.}

After setting $\lambda^\star = 2.4837$, the posterior mean is $c_0=0.309$ (median $0.299$, 90\% CI $[0.122,\ 0.481]$) and $\kappa=0.1501$ (median $0.1506$, 90\% CI $[0.1329,\ 0.1658])$. The wall time is $358\,\mathrm{s}$ in total for $4$ chains. We obtain $\hat R_{c_0} = 1.011$ and $\hat R_{\kappa} = 1.010$ without divergences. The respective effective sample sizes for $c_0$ and $\kappa$ are $351$ and $322$.
Overall, these results suggest that the proposed method performs well in such regimes without convergence issues. 

\subsection{Experiment C: Misspecified Gaussian pulse model with errors-in-variables}
\label{sec:expC}
\subsubsection{Experiment description.}
We study an errors-in-variables signal calibration problem. The goal is to evaluate whether our Sinkhorn-based method can outperform a likelihood baseline. This experiment highlights the geometric robustness of OT losses. We define a Gaussian pulse model with added timing jitter.
\paragraph{Model, data and choice of priors.}
Let $u_i\in[0,1]$ be latent time locations,

\begin{equation}
\xi_i \overset{\text{i.i.d.}}{\sim} \mathcal{N}\!\left(0,\ \texttt{jitter\_scale}^2\right),\qquad
t_i^{\mathrm{obs}} = u_i + \xi_i,\qquad
y_i^{\mathrm{obs}} = f(u_i;\theta^\star) + \zeta_i,
\end{equation}

\begin{equation}
f(t;\theta) \;=\; \mathrm{baseline} + \mathrm{amp}\cdot \exp\!\left(-\tfrac12 \big((t-t_0)/\mathrm{width}\big)^2\right).
\end{equation}
Here  $\xi_i$ corresponds to the timing jitter. The true parameter values are set as follows: $\theta^\star=(\mathrm{baseline}^\star=0,\
\mathrm{amp}^\star=1,\ t_0^\star=0.5,\ \mathrm{width}^\star=0.05,\  \sigma^\star=0.02)$,
with $\texttt{jitter\_scale}=0.05$. The vertical noise is defined as $\zeta_i \overset{\text{i.i.d.}}{\sim} \mathcal{N}(0, \sigma^2)$.

The resulting observations are the point cloud $\{(t_i^{\mathrm{obs}},y_i^{\mathrm{obs}})\}_{i=1}^N$. We define the following priors: 
$\mathrm{baseline}\sim \mathcal{N}(0,1)$, $\mathrm{amp}\sim \mathrm{HalfNormal}(2)$, $t_0\sim \mathrm{Beta}(2,2)$, $\mathrm{width}\sim \mathrm{HalfNormal}(0.2)$ and $\sigma\sim\mathrm{HalfNormal}(0.2)$.
As a likelihood baseline, we define a vertical-noise regression model that treats $t^{\mathrm{obs}}$ as exact:
\begin{equation}
  y_i^{\mathrm{obs}} \mid t_i^{\mathrm{obs}},\theta \sim \mathcal{N}\!\big(f(t_i^{\mathrm{obs}};\theta),\sigma^2\big).
\end{equation}
% IL FAUT DIRE POURQUOI ON D2FINIT UN SIGMA ICI
\newcommand{\expBtwofig}{\figpath/expB2}

\paragraph{OT loss settings and HMC/NUTS selected calibration parameters.}
We use the same configuration as \hyperref[sec:expA_normal]{experiment A}. The resolution yields $\varepsilon=0.00327$.

\subsubsection{Results and interpretations.} 
\begin{figure}[t]
\centering
\scalebox{0.33}{\input{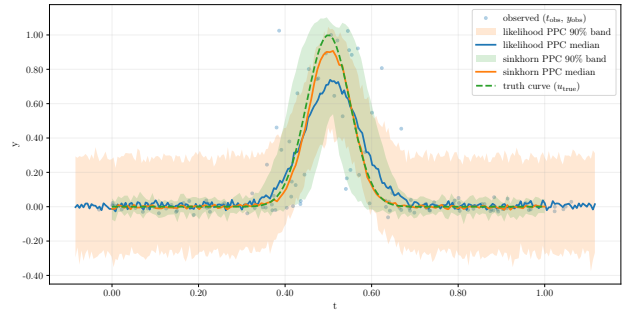}}
\caption{Experiment C: Observed point cloud and posterior predictive overlays.}
\label{fig:expB2_ppc}
\end{figure}

We start by conducting a likelihood-based calibration while disabling the timing jitter, and obtain sensible results\footnote{Posterior (parameter, mean, 5\%, 95\%, $\hat{R}$):
$(\mathrm{base.}, -0.00218,\,-0.00556,\,0.00129,$ $1.00000)$;
$(\mathrm{amp}, 1.0073,\,0.9941,\,1.0201,\,1.0025)$;
$(\sigma, 0.0215,\,0.0192,\,0.0234,\,1.0017)$;
$(t_0, 0.5004,\,0.4997,\,0.5012,\,0.9985)$; $(\mathrm{width}, 0.0498,\,0.0491,\,0.0505,\,1.0064)$.}. Adding the timing jitter, we obtain the following results presented in Table \ref{tab:expC_normal_summary_compact2}.
\begin{table}[t]
%AJouter une précision
\caption{Experiment C: results of balanced Sinkhorn versus baseline.}
\label{tab:expC_normal_summary_compact2}
\centering
\small
\setlength{\tabcolsep}{3pt}
\renewcommand{\arraystretch}{1.1}
\begin{tabular}{llcccc}
\toprule
\multicolumn{1}{c}{Method} &
\multicolumn{1}{c}{Parameter} &
\multicolumn{1}{c}{Truth} &
\multicolumn{1}{c}{mean (sd) [5\%, 95\%]} &
\multicolumn{1}{c}{$n_{\text{eff}}$($4\times300$)} &
\multicolumn{1}{c}{$\hat R$} \\
\midrule
\multirow{5}{*}{\parbox[c]{2.7cm}{\centering\textit{Likelihood\\baseline}}}
 & amp      & 1.00 & 0.724 (0.059) [0.624, 0.818] & 975  & 1.001 \\
 & baseline & 0.00 & 0.006 (0.019) [-0.024, 0.038] & 916  & 1.001 \\
 & sigma    & 0.02 & 0.175 (0.011) [0.157, 0.192] & 1150 & 1.002 \\
 & $t_0$       & 0.50 & 0.510 (0.005) [0.500, 0.518] & 1086 & 1.001 \\
 & width    & 0.05 & 0.065 (0.007) [0.053, 0.075] & 705  & 1.001 \\
\midrule
\multirow{5}{*}{\parbox[c]{2.7cm}{\centering\textit{Balanced\\Sinkhorn}}}
 & amp      & 1.00 & 1.010 (0.079) [0.881, 1.142] & 807 & 1.000 \\
 & baseline & 0.00 & -0.002 (0.019) [-0.032, 0.031] & 817 & 0.998 \\
 & sigma    & 0.02 & 0.022 (0.015) [$1.7\times10^{-5}$, 0.043] & 613 & 1.007 \\
 & $t_0$       & 0.50 & 0.508 (0.028) [0.468, 0.555] & 935 & 1.000 \\
 & width    & 0.05 & 0.048 (0.010) [0.031, 0.062] & 714 & 1.000 \\
\bottomrule
\end{tabular}
\end{table}
They indicate that the likelihood-based calibration is strongly impacted by the timing jitter. While $\hat{R}$ and $n_{\text{eff}}$ remain satisfactory, we see that $\sigma$ is strongly inflated ($0.175 > 0.02$) and the amplitude is underestimated. Both these parameters end up outside of the $90\%$ credible interval. Meanwhile, the balanced Sinkhorn calibration results show no issues, with all true parameters in the $90\%$ credible interval. The PPC bands visualization in Figure \ref{fig:expB2_ppc} also confirms this superiority. We observe no divergences in either of the calibration processes.
These results suggest that our Sinkhorn-based HMC method is particularly robust in certain errors-in-variables situations, compared to likelihood-based alternatives. 

\subsection{Experiment D: CIFAR-10 image patch alignment under perturbations}
\label{sec:expD}
\subsubsection{Experiment description.} In this experiment, we test our method on an image dataset\footnote{CIFAR-10, available at \url{www.cs.toronto.edu/~kriz/cifar.html}.} with a deterministic simulator. OT discrepancies are frequently used for computer vision tasks like image generation \cite{wassersteingan,genevay_learning_2017} or object detection \cite{wassOD}. We design an experiment in which the practical difference between balanced and unbalanced OT losses is clear. We also consider a sweep over loss scales.

\begin{figure}[t]
\centering
\scalebox{0.35}{\import{}{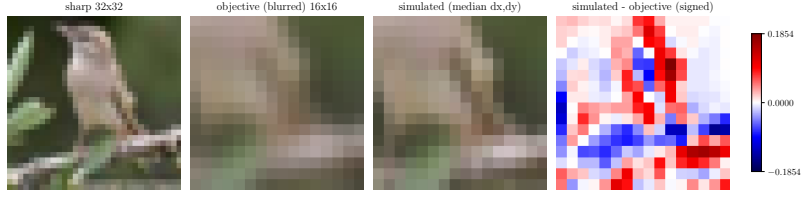}}
\caption{Objective alignment example and average signed channel difference.}
\label{fig:registration}
\end{figure}

\paragraph{Model, data and choice of priors.}
We use the images from the dataset without any selection based on classes or any other criteria. We sample a $2D$ offset (based on the prior described below) which defines the target offset. The sampled offset $(dx,dy)$ is different for each image (i.i.d.). We then apply this offset to the base $32\times32$ image center. Using this computed point as a new center, we select the square $16\times16$ window around it. Bilinear interpolation is used because the offset distribution is continuous. We apply a Gaussian blur ($\text{kernel size} = 3\times3, \Sigma = 1.5$) on it to get our final objective window, see Figure \ref{fig:registration}.

In our experiment, we set $\sigma=4$ and $d_{\max} = 8$ meaning that the objective window is always contained within the original $32\times32$ image. 

\begin{equation}
    d_x^\star \sim \mathcal{N}_{\text{trunc}}(0,\sigma^2;\,-d_{\max},d_{\max}),
    \qquad
    d_y^\star \sim \mathcal{N}_{\text{trunc}}(0,\sigma^2;\,-d_{\max},d_{\max}),
\end{equation}
\noindent where $\mathcal{N}_{\text{trunc}}$ is the truncated normal distribution.
The calibration task is to explore the original $32\times32$ sharp image with a $16\times16$ field of view. Any candidate window is computed using the original image, the given offset $\theta=(d_x,d_y)$ and bilinear interpolation. Sinkhorn is applied between the objective window (target offset with added blur) and the candidate window (both $16\times16$). Bilinear interpolation preserves differentiability, which explains why it's commonly used in such situations \cite{SPT}. The loss is averaged over the ${R,G,B}$ color channels. Introducing a luminance offset $\psi = 10^{-6}$ to improve numerical stability, we get the following loss computation scheme for the balanced, normalized settings:

%\begin{equation}
%s_c \;=\; \frac{1}{32^2}\sum_{p\in\{0,\dots,31\}^2} I_{\text{sharp}}^{(c)}(p)
%\qquad (c\in\{R,G,B\}),
%\end{equation}

%\begin{equation}
%a_i^{(c)}(\theta)=\frac{X^{(c)}(\theta)(x_i)+\psi}{s_c},
%\qquad
%b_i^{(c)}=\frac{Y^{(c)}(x_i)+\psi}{s_c}.
%\end{equation}

%\begin{equation}
%\tilde a_i^{(c)}(\theta)=\frac{a_i^{(c)}(\theta)}{\sum_{k=1}^n a_k^{(c)}(\theta)},
%\qquad
%\tilde b_i^{(c)}=\frac{b_i^{(c)}}{\sum_{k=1}^n b_k^{(c)}}.
%\end{equation}
%\begin{equation}
%\tilde\alpha_\theta^{(c)}=\sum_{i=1}^n \tilde a_i^{(c)}(\theta)\,\delta_{x_i},
%\qquad
%\tilde\beta^{(c)}=\sum_{i=1}^n \tilde b_i^{(c)}\,\delta_{x_i}.
%\end{equation}
%\begin{equation}
%\mathcal{L}(\theta)=\frac{1}{3}\sum_{c\in\{R,G,B\}}
%S_{\varepsilon}\!\big(\tilde\alpha_\theta^{(c)},\tilde\beta^{(c)}\big).
%\end{equation}

%For the unbalanced settings, we do not normalize the image.

%\begin{equation}
%\alpha_\theta^{(c)}=\sum_{i=1}^n a_i^{(c)}(\theta)\,\delta_{x_i},
%\qquad
%\beta^{(c)}=\sum_{i=1}^n b_i^{(c)}\,\delta_{x_i}.
%\end{equation}
%\begin{equation}
%\mathcal{L}^{\mathrm{U}}(\theta)=\frac{1}{3}\sum_{c\in\{R,G,B\}}
%S_{\varepsilon,\rho}^{\mathrm{U}}\!\big(\alpha_\theta^{(c)},\beta^{(c)}\big).
%\end{equation}

%\begin{equation}
%s_c \;=\; \frac{1}{32^2}\sum_{p\in\{0,\dots,31\}^2} I_{\text{sharp}}^{(c)}(p)
%\qquad (c\in\{R,G,B\}),
%\end{equation}

\begin{equation}
s_c \;=\; \frac{1}{32^2}\sum_{p\in\{0,\dots,31\}^2} I_{\text{sharp}}^{(c)}(p)
\qquad (c\in\{R,G,B\}).
\end{equation}

\begin{equation}
a_i^{(c)}(\theta)=\frac{I_\theta^{(c)}(x_i)+\psi}{s_c},
\quad
b_i^{(c)}=\frac{I_{\mathrm{obs}}^{(c)}(x_i)+\psi}{s_c},
\quad
\{x_i\}_{i=1}^{n=16^2}\subset\mathbb{R}^2.
\end{equation}

\begin{equation}
\tilde a_i^{(c)}(\theta)=\frac{a_i^{(c)}(\theta)}{\sum_{k=1}^n a_k^{(c)}(\theta)},
\qquad
\tilde b_i^{(c)}=\frac{b_i^{(c)}}{\sum_{k=1}^n b_k^{(c)}}.
\end{equation}

\begin{equation}
\tilde\mu_\theta^{(c)}=\sum_{i=1}^n \tilde a_i^{(c)}(\theta)\,\delta_{x_i},
\qquad
\tilde\nu^{(c)}=\sum_{i=1}^n \tilde b_i^{(c)}\,\delta_{x_i}.
\end{equation}
\begin{equation}
\mathcal{L}(\theta)=\frac{1}{3}\sum_{c\in\{R,G,B\}}
S_{\varepsilon}\!\big(\tilde\mu_\theta^{(c)},\tilde\nu^{(c)}\big).
\end{equation}

For the unbalanced settings, we do not normalize the image.

\begin{equation}
\mu_\theta^{(c)}=\sum_{i=1}^n a_i^{(c)}(\theta)\,\delta_{x_i},
\qquad
\nu^{(c)}=\sum_{i=1}^n b_i^{(c)}\,\delta_{x_i}.
\end{equation}
\begin{equation}
\mathcal{L}^{\mathrm{U}}(\theta)=\frac{1}{3}\sum_{c\in\{R,G,B\}}
S_{\varepsilon,\rho}^{\mathrm{U}}\!\big(\mu_\theta^{(c)},\nu^{(c)}\big).
\end{equation}

\paragraph{OT loss settings.}
Sinkhorn iterations are set to $64$, and CRN is disabled. We set $\alpha = 1$, hence $\varepsilon = 1$, since the natural spatial resolution is one pixel. We sweep through loss scales (0.06, 0.12, 0.25, 0.5, 1, 2, 4, 8) in both the balanced/normalized and unbalanced regimes. In the latter, we set $\rho=10$, so $\sqrt{10}$ pixels is the approximate scale at which mass variation competes with transport.%In total, we get $704$ configurations.
\begin{figure}[t]
\centering
\scalebox{0.36}{\import{}{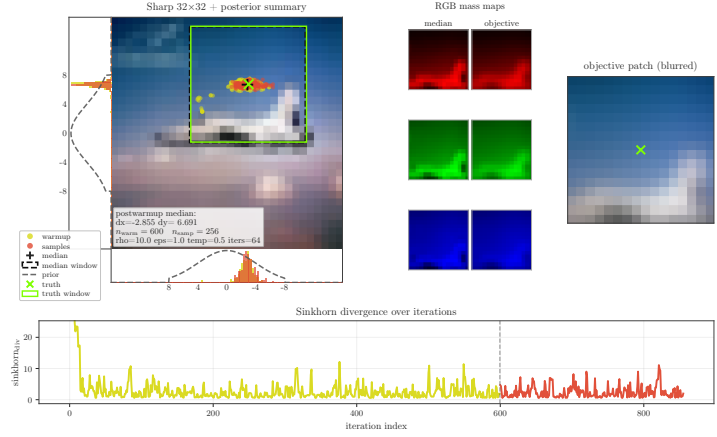}}
\caption{Warmup, sampling process, and posterior median results on a held out image (not included in the loss scale calibration set).}
\label{fig:panel}
\end{figure}
\paragraph{HMC/NUTS selected calibration parameters.}
We sample $4$ chains each with $600$ warmup and $256$ retained samples. The target acceptance rate is set to $0.75$.
\subsubsection{Results and interpretations.}
Here, we compute the coverage $\mathrm{cov}_{2D}$ and corresponding area $A_{2D}$ from a Gaussian credible ellipse approximation. 
The unbalanced approach, by avoiding normalization, can account for both local transport and mismatch in total mass. In comparison, the balanced approach is not able to produce satisfactory results as the accuracy is much worse, with coverage that rapidly deteriorates with increasing loss scale.
After selecting $\lambda^\star=0.50$ with the loss scale sweep, we check the results on $32$ new images. We see that the performance remains good although the convergence diagnostics are not uniformly ideal, see Table \ref{tab:expD}. These limitations are partly due to strong heterogeneity in curvature. Without a higher number of samples, large steps between stability regions may not have enough time to occur. That said, the posterior sample distributions nicely depict the uncertainty geometry, see Figure \ref{fig:panel}. 

The average wall time over the 32 held-out images is $74.4\,\mathrm{s}$ per chain.

\begin{table}[t]
\caption{Experiment D results, $\dagger$: proportion of cases with $\|\hat\theta_{0.5}-\theta^\star\|_2>4$.}
\label{tab:expD}
\centering
\small
\setlength{\tabcolsep}{4pt}
\begin{tabular}{lccccccc}
\toprule 
\shortstack[c]{$\lambda$} &
\shortstack[c]{$\widehat{\mathrm{cov}}_{2D}$\\(0.95)} &
\shortstack[c]{$\mathrm{med}$\\$\|\hat\theta_{0.5}-\theta^\star\|_2$} &
\shortstack[c]{$\mathrm{P90}$\\$\|\hat\theta_{0.5}-\theta^\star\|_2$} &
\shortstack[c]{$\dagger$\\(\%)} &
\shortstack[c]{$\mathrm{med}$\\$A_{2D}(0.90)$} &
\shortstack[c]{div.\\(\%)} &
\shortstack[c]{$\hat R_{\max}\le 1.1$\\(\%)} \\
\midrule
\multicolumn{8}{c}{Balanced regime loss scale sweep (first 32 images), relevant $\lambda$ values subset} \\
\midrule
%0.06 & 0.969 & 0.220 & 0.876 & 3.9 & 6.73 & 0.0 & 84.4 \\
%0.12 & 0.914 & 0.130 & 0.519 & 8.6 & 3.18 & 0.0 & 78.1 \\
0.25 & 0.898 & 0.108 & 1.145 & 9.4 & 1.39 & 0.8 & 68.8 \\
0.50 & 0.859 & 0.097 & 8.680 & 14.1 & 0.67 & 0.8 & 56.2 \\
1.00 & 0.820 & 0.089 & 9.476 & 14.8 & 0.33 & 0.0 & 62.5 \\
%1.50 & 0.773 & 0.099 & 10.064 & 19.5 & 0.23 & 0.0 & 46.9 \\
2.00 & 0.750 & 0.110 & 10.988 & 20.3 & 0.17 & 0.8 & 46.9 \\
%3.00 & 0.742 & 0.103 & 9.955 & 19.5 & 0.12 & 1.6 & 43.8 \\
%4.00 & 0.656 & 0.119 & 9.193 & 21.1 & 0.08 & 0.0 & 53.1 \\
%6.00 & 0.547 & 0.125 & 10.829 & 28.1 & 0.05 & 0.8 & 31.2 \\
%8.00 & 0.508 & 0.126 & 11.324 & 25.8 & 0.04 & 0.8 & 34.4 \\
\midrule
\multicolumn{8}{c}{Unbalanced regime loss scale sweep (first 32 images), relevant $\lambda$ values subset} \\
\midrule
%0.06 & 0.992 & 0.504 & 2.674 & 3.1 & 24.79 & 0.0 & 96.9 \\
%0.12 & 0.977 & 0.271 & 1.452 & 2.3 & 10.80 & 0.0 & 90.6 \\
0.25 & 0.984 & 0.109 & 0.609 & 1.6 & 4.52 & 0.8 & 93.8 \\
0.50 & 0.953 & 0.091 & 0.290 & 4.7 & 2.04 & 1.6 & 90.6 \\
1.00 & 0.891 & 0.080 & 7.569 & 10.9 & 0.99 & 3.9 & 71.9 \\
2.00 & 0.906 & 0.083 & 0.343 & 9.4 & 0.47 & 0.8 & 75.0 \\
%4.00 & 0.875 & 0.091 & 4.421 & 10.9 & 0.23 & 1.6 & 65.6 \\
%8.00 & 0.781 & 0.092 & 7.942 & 14.8 & 0.11 & 0.8 & 59.4 \\
\midrule
\multicolumn{8}{c}{Held-out test set (32 new images, selected \(\lambda^\star=0.50\), unbalanced regime)} \\
\midrule
0.50 & 0.961 & 0.127 & 0.424 & 3.9 & 3.42 & 0.0 & 87.5 \\
\bottomrule
\end{tabular}
\end{table}

% Requires: \usepackage{booktabs}

%\section{Discussion}
\section{Conclusion and Future Work}
In this work, we introduced Sinkhorn divergences as Generalized Bayes losses for posterior inference, leveraging the geometric properties of optimal transport losses while staying compatible with HMC methods. This approach is especially relevant when likelihood functions are misspecified or poorly suited, for instance in the context of errors-in-variables.

In addition to the theoretical formulation, we proposed a practical algorithmic framework that can handle both balanced and unbalanced divergences. We define a surrogate deterministic loss that fixes the randomness of stochastic simulators, broadening applicability. Across multiple empirical experiments, we showed that the method (1) yields sensible results in comparison with likelihood-based models; (2) provides robust calibration when likelihood methods are inappropriate; (3) shows validity under misspecification and horizontal noise; (4) highlights the situational effectiveness of unbalanced regimes. The diversity of considered datasets further suggests that the method is robust enough to be applicable across several useful application domains. Our practical heuristics for loss-parameter selection lead to stable and meaningful calibration results.

In future work, the effects of a low Sinkhorn iteration budget could be theoretically and empirically studied, as they may induce additional regularization that may be beneficial in some settings. It would also be interesting to apply the method to econometric model calibration.

\begin{credits}
\subsubsection{\ackname} The authors thank colleagues at LIST3N and QAD Inc. for their support. This work is supported by an ANRT CIFRE industrial Ph.D. fellowship.
\end{credits}
%% Note that this preceding line implies that you store your BibTeX references in a file called 'mybibliography.bib'. If you instead store your references in a file with a different name, for instance 'references.bib', the preceding line should read '\bibliography{references}'. Whatever you do, DO NOT put the file name extension .bib inside the \bibliography command; this will trip up LaTeX compilers. 
%
% If you do not want to use BibTeX, you can also type up the bibliography exactly as you see fit, using the following structure:
\bibliographystyle{splncs04}
\bibliography{references}

\end{document}